\documentstyle[11pt]{article}

\begin{document}
\begin{flushright}
\today\\
\end{flushright}
\vspace{1 cm}
\begin{center}
{\Large \bf Water temperature constraint on Sonoluminescence}\\
\vspace{1 cm}
{\it \large Sohrab Rahvar \footnote{e-mail:rahvar@physics.sharif.ac.ir} }\\
\vspace{3 mm}
{\it\large Atomic Energy Organization of Iran, Bonab Research Center,P.O.Box 
 11365--8486, Tehran,Iran}\\
{\it \large Institute for Studies in Theoretical Physics and Mathematics,
P.O.Box 19395--5531, Tehran, Iran}\\
\end{center}
\begin{abstract}
It is proposed that shock wave dynamics within the gas of
a small bubble explain sonoluminescence, the emission of
visible radiation. As the bubble radius oscillates, shock waves
develop from spherical sound waves created inside the gas
bubble. As any such shock propagates toward the center, it
strengthens and, upon convergence and subsequent reflection, temperature
of gas inside bubble increases dramatically in such a way
that it can produce plasma. Since main radiation product in exploding
epoch, nonadiabatic condition for imploding shock wave cool plasma
and cause exploding shock wave can not sufficiently rise temperature
to produce radiation. In this work we compare cooling time for plasma by
bermsstrahlung radiation with collapsing time for the imploding shock wave
. We find a constraint on radius of bubble with respect to temperature of 
water. This constraint condition
explains experimental results as to, why the cold water is fine for SL.
\end{abstract}
\newpage
\section{Introduction}
Sonoluminescence, the phenomenon of emission of light from small bubble
occurring during ultrasonic excitation, has been known for more than half
a century. In this phenomenon an intense standing wave increases 
the pulsation of a bubble of 
gas trapped at a velocity node to attain sufficient amplitude so as to emit 
picosecond flash of light.
The analysis of the dynamics of a small bubble or cavity in a fluid dates back
to the work of Lord Rayleigh [1] at the beginning of
this century. A large number of publications followed in subsequent decades,
including the studies of oscillating bubbles by Plesset [2], Eller $\&$
Crum , Flynn[3], Lauterborn[4], Prosperetti[5], and many others. Experiments indicate 
that the collapse is
remarkably spherical, and that water is the best fluid for SL. Some noble gas is
essential for stable SL, and that the light intensity increases as ambient 
temperature is lowered. The theory of light emitting mechanism is still open, but 
the traditional scenario is supersonic bubble collapse launching an imploding 
shock wave which ionizes the bubble contains so as cause it to emit.
Bremsstrahlung radiation is the best theory for SL phenomenon. In the theory 
of radiation by Bremsstrahlung as the bubble collapses, produced shock wave
ionized the gas and cause it to radiate. The imploding shock wave concentrated 
at the center of bubble, through this
concentration, shock wave warmed the gas until
will receive to the center of bubble and explode, through this explosion
gas which was compressed just behind the imploding shock front now finds 
itself in front of the shock front again. As the shock passes through 
these particles a second time, there is another burst of heating and
maximum temperature reached by the exploding shock wave [6].
The mean idea of our work is that, if in the imploding regime the 
plasma cools by radiation then in the exploding epoch, shock front can 
not warm particles again to produce intensive radiation. 
Thus by comparing cooling time with dynamical time for collapsing,
we can say that if cooling time is less than collapsing time then plasma 
inside bubble will cool and the exploding shock wave can not warm 
this gas again. This constraint
gives us some useful parameter for SL, therefore, we can explain results of 
some experiments such as why the light intensity
increases as ambient temperature is lowered[7].
\section{Collapse Mechanism and radiation}
Explanation of the light-emitting mechanism of SL naturally seeks to
interpret the featureless spectrum in terms of emission from a hot spot, for
example black body radiation, if the radiation and matter are near to 
equilibrium, or Bremsstrahlung from accelerating unbound
electrons if the light-emitting region is hot enough to be ionized yet
sufficiently rarefied so as to be transparent to radiation[8]. 
The shock wave model [9,10,6,11] provides an extra stage of
energy focusing by assuming that supersonic inward collapse of the bubble
wall launches a shock wave into the bubble's interior. This shock can run
through the already compressed gas inside the bubble, increasing its
amplitude and speed as it focuses towards the origin. There is now a surface
of radius $R_s$ (the radius of the shock front), within the bubble of 
radius $R$. The similarity solution is obtained by assuming that the 
shock radius takes this form :
\begin {equation}
R_s=A(-t)^\alpha,
\end {equation}
(Guderley 1942)
[12] where time is measured from the time of the convergence
of the shock, and A is the "launch" condition of the shock, which couples
the shock to the bubble motion. The similarity solution
yields an exponent $\alpha $ of 0.72 in air and 0.69 for noble 
gases. Since the
exponent is less that unity, the Mach number of shock as goes to the origin
approaches infinity. The bubble wall is collapsing at the speed of sound when
it is passing through its ambient radius. In this case (Barber et al)[6] showed 
that:
\begin {equation}
R_s=R_0(-\frac t{t_0})^\alpha,
\end {equation}
where
\begin {equation}
t_0=\alpha \frac{R_0}{c_0},
\end {equation}
In terms of Mach number
\begin {equation}
M=\left| \frac{t_0}{t}\right| ^{1-\alpha }.
\end {equation}
one can define dynamical time or collapsing time according to:
\begin {eqnarray}
t_{dy}&=&\frac{R(t)}{\frac{dR(t)}{dt}},\\
t_{dy}&=&\frac{\alpha R_0}{c_0}(\frac {1}{M})^{\frac{1}{1-\alpha }}.
\end {eqnarray}
where $t_{dy}$ is the time scale for the shock to have a radius smaller 
than $R_s$.
As the shock wave approaches the center the temperature immediately behind
the imploding shock front increases by this factor $\frac T{T_0}=M^2$.
When the shock wave converges to the origin it explodes from the origin 
with the same similarity solution. Thus the gas which was compressed just behind 
imploding shock front now finds itself in the front of the shock
front again. As the shock passes through these particles a second time, 
there is another burst of heating and the maximum temperatures reached by the
exploding shock wave now increases by the factor $\frac T{T_0}=M^4$[7].
\section{Cooling Condition for Radiation}
The temperature increase inside the bubble by shock wave ionizes the region.
The free electrons released by the heating will accelerate and radiate light as
they collide with the ions. The Bremsstrahlung radiation generated per second per
volume is equal to :
\begin {equation}
\frac {dE}{dt dv}=\frac{16z^{2}e^{6}n^{2}}{3m^{2}c^{3}h}(\frac{2mkT}{\pi})^{1/2}g_{B}\approx
1.5\times10^{-27}n^{2}T^{1/2}g_{B}.
\end {equation}
where $e$ and $m$ are the electron charge and mass, $n$ the density of free
electrons and ions, $c$ the speed of light, and $g_{B}$ the
Gount factor. In the temperature of T, the energy density of this plasma 
is equal to:
\begin {equation}
\frac{dE}{dv}=\frac{3}{2}nkT.
\end {equation}
So we can introduce cooling time for this plasma
\begin {eqnarray}
t_{cool}&=&\frac{\frac{dE}{dv}}{\frac{dE}{dt dv}}.\\
t_{cool}&=&\frac{9m^{2}c^{3}h}{32z^{2}e^{6}}(\frac{\pi}{3mk})^{1/2}n^{-1}T^{1/2}.\\
\approx \frac{0.26}{g_{B}}\times 10^{12}n^{-1}T^{1/2}.
\end {eqnarray}
As the bubble collapses
the value of $n$ changes with the radius, by using the dynamical equation of
shock wave we can interpret the above density in terms of initial density 
and radius of gas in the bubble as:
\begin {equation}
n=n_{0}(\frac {R_{0}}{R_{s}})^{3}.
\end {equation}
By using dynamical equation for shock wave, cooling time scale 
can be obtained in terms of Mach number
\begin {equation}
t_{cool}=\frac{0.26}{g_{B}} \frac{10^{12}}{z^{2}n_{0}}(\frac{1}{M})^{\frac {3\alpha}{1-\alpha}}T^{1/2}.
\end {equation}
Now we compare this time scale with dynamical time scale for the shock wave.
If $t_{cool}>t_{dy}$ then for the imploding shock wave, as the shock wave 
implodes toward the origin, plasma cannot cool thus adiabatic condition
holds. So in the explosion epoch temperature can rise sufficiently to produce
an extensive radiation. Now if $t_{cool}<t_{dy}$, then as
the imploding shock wave implodes towards the center of bubble by the radiation
mechanism, plasma cools and temperature can not rise in such a way that 
exploding shock can produce an intensive radiation.
As mentioned before, temperature behind the shock wave rises by a factor 
of $\frac {T}{T_{R}}=M^{2}$, where $T(R)$ is the 
temperature of the bubble when its radius is $R$. For the constraint 
$t_{cool}>t_{dyn}$, temperature of gas
inside the bubble rises with an adiabatic compressing:
\begin {equation}
T(R)=T_{0}(\frac {R_{0}}{R})^{3(\gamma-1)},
\end {equation}
where $ T_{0} $ is the ambient temperature of bubble and 
$R_{0}$ is the size of the bubble when it is in a static 
mechanical equilibrium. By using shock wave dynamics, the 
above equation can be obtained in terms of Mach number, so we have:
\begin {equation}
T(R)=T_{0}M^{\frac {3}{2}\frac {\alpha}{1-\alpha}(\gamma-1)}.
\end {equation}
By substituting Eqs.(6 \& 13) in the adiabatic condition for imploding shock
wave, constraint condition obtain in the term of ambient temperature, Mach
number, initial density and initial radius and constants of gas
\begin {equation}
R_{0}<\frac{0.26}{g_{B}} \frac{c_{0}10^{12}}{\alpha zn_{0}}{M}^{\frac {3\alpha\gamma-11\alpha+4}
{2(1-\alpha)}}T_{0}^{1/2},
\end {equation}
where $c_{0}$ is the velocity of sound and $n_{0}$ is the initial density
of gas inside the bubble. From experiment we have $n_{0}\approx
4.16\times 10^{19}\frac{par}{cm^3}$ and $g_B=1.2$. Measurements 
indicate that the bubble wall is collapsing at
more than 4 times the ambient speed of sound in air[13].
So the above inequality can be written this way:
\begin {equation}
R_{0}<0.258\times 10^{-4}{T_{0}}^{1/2}.
\end {equation}
As we said temperature rises by two processes:\\
i. By shock wave $T=M^{2}T(R)$.\\
ii. By adiabatic collapsing of bubble$T(R)=M^{(\gamma-1)
(\frac{3\alpha}{1-\alpha})}T_{0}$.\\
The factor of increase in temperature in the above processes is $T=M^{2}M^{(\gamma-1)
(\frac{3\alpha}{1-\alpha})}T_{0}$. For air, we have
$T=1144\times T_{0}$ where $T_{0}$ is the ambient temperature, If we consider room
temperature for ambient temperature, the temperature of bubble 
rises up to $10^{5}$. In the Bremsstrahlung radiation we considered all 
of our gases to be ionized. Verification of that statement can be 
obtained by Saha's equation 
\begin {equation}
\frac{q^{2}}{1-q}=2.4\times10^{21}T^{3/2}exp({-\frac{\chi}{kT}}){\frac{1}{n}}
\end {equation}
where $q=\frac {n_{e}}{n}$ is the degree of ionization, $\chi$ is the 
ionization potential, and the pre factor $T$ is given 
in Kelvin. If we put the given parameters for SL 
in Saha's equation for air, we find that $n_{e}=n_{i}=n$. This is 
the reason for considering completely ionized gas Bremsstrahlung formula.
The inequality curve of Eq.(17) gives us a constraint for producing radiation 
in SL (Fig). On the other hand Bradley et al (1994) show the extreme sensitivity
of SL to external parameters such as the water temperature. They 
show that as the water temperature decreases from $40^\circ$ to $1^\circ$, the 
intensity of light emission increases by a factor of over $ 200$ . 
At about $0^\circ$ the purple light emitted by the bubble is so bright that one can see it by
an unaided eye, but at $40^\circ$ the SL is barely visible even in a darkened
room. According to Bradley et al (1994)
by light scattering technique initial radius of bubble obtained respect to
ambient temperature. Comparing constraint curve for SL with the experimental 
relation between ambient radius of bubble and temperature (Fig), we see 
that above about $25^\circ$ constraint for SL breaks down and confirms 
the experiment results of Bradley et al (1994).
\section{Acknowledgment}
I would like to thank Dr.Rasool Sadighi and Dr.Kamal Seied yaghobi for usefull discussion and comments.
 
\newpage
\section{Figure  Caption}
Bar lines indicated the experimental measurement between ambient temperature
and initial radius of bubble. On the other hand there is a constraint line
form Eq.(17), in such a way that below this line flashes of light can
produce in SL. About $25^{\circ }$ experimental bars cross our line and
causes to SL break downs. This confirms the results of Bradley et at (1994)
\end{document}